\documentclass[11pt,a4paper]{article}

\usepackage[utf8]{inputenc}
\usepackage[british]{babel}
\usepackage[margin=20mm,top=20mm,bottom=20mm]{geometry}
\usepackage[pdftex]{hyperref}
\hypersetup{pdfauthor={J. Cuenca, P. Göransson, L. De Ryck, T. Lähivaara},pdftitle={Deterministic and statistical characterisation of poroelastic media}}

\usepackage{amsmath,xcolor,graphicx,booktabs,tikz,siunitx,subfig}

\newcommand\ig[2]{\includegraphics[width=#1\linewidth]{#2}}
\newcommand\igh[2]{\includegraphics[height=#1]{#2}}

\newcommand{\Rho}{\mathrm{P}}
\newcommand\fobj{f_\text{obj}}
\newcommand\xOdet{\widetilde\mathbf{x}_0^\text{(det)}}
\newcommand\nit{$n_\text{it}$}

\newcommand{\loadA}{(A)}
\newcommand{\loadB}{(B)}

\newcommand\gmeas{g^{\text{meas}}}
\newcommand\gmeasbold{\mathbf{g}^{\text{meas}}}
\newcommand\alphameas{\alpha^{\text{meas}}}

\newcommand{\blackline}{\raisebox{3pt}{\tikz{\draw[-,black,solid,line width = 1pt](0,0) -- (7mm,0);}}}
\newcommand{\credline}{\raisebox{3pt}{\tikz{\draw[-,black!50,line width = .5pt, dash pattern = on 1.5pt off 1.2pt](0,0) -- (7mm,0);}}}
\definecolor{grey}{rgb}{.8,.8,.8}
\newcommand{\greydottedline}{\raisebox{3pt}{\tikz{
      \draw[black!50,solid,line width = .3pt](0,0) -- (7mm,0);
      \draw[black!50,line width=1.5pt, line cap=round, dash pattern=on .7pt off 2pt] (0,0) -- (7mm,0);
    }}}
\definecolor{blueline}{rgb}{0 .45 .74}
\newcommand{\blueline}{\raisebox{2pt}{\tikz{\draw[-,blueline,solid,line width = 1.5pt](0,0) -- (7mm,0);}}}
\newcommand{\blackdottedline}{\raisebox{2pt}{\tikz[black]{
      \draw[line width=.3pt](0,0) -- (7mm,0);
      \draw[line width=2pt, line cap=round, dash pattern=on .3pt off 3.5pt] (0,0) -- (7mm,0);
    }}}
\definecolor{lightteal}{rgb}{0 .74 .45}
\newcommand{\tealline}{\raisebox{1pt}{\tikz[lightteal,line width=1.4pt, line cap=square]{
      \draw[dash pattern=on 1.4pt off 1.4pt] (0,0) -- (6.4mm,0);
      \draw[line width = 3pt, line cap = round](3.2mm,0) -- (3.2mm,0);
      \draw (6.8mm,-.6mm) -- (6.8mm,.6mm);
      \draw (-.4mm,-.6mm) -- (-.4mm,.6mm);
    }}}
\definecolor{thickteal}{rgb}{.5 .75 .84}
\newcommand{\thicktealline}{\raisebox{1pt}{\tikz[thickteal,line width=3pt, line cap=square]{\draw(0,0)--(6.4mm,0);}}}

\definecolor{CMcircle}{rgb}{.2 .2 .2}
\definecolor{MAPcross}{rgb}{0 0 0}
\newcommand{\CMcircle}{\raisebox{1pt}{\tikz{\draw[-,black,solid,line width = .7pt](0,0) circle (1.5pt);}}}
\newcommand{\MAPcross}{\raisebox{1pt}{\tikz[-,black,solid,line width = .7pt]{\draw(0,0)--(3pt,3pt); \draw(0,3pt)--(3pt,0);}}}

\newcommand{\ra}{r_\text{a}}
\newcommand{\la}{l_\text{a}}
\newcommand{\lb}{l_\text{b}}
\newcommand{\lbO}{\bar{l}_\text{b}}
\newcommand{\lc}{l_\text{c}}
\newcommand{\lcO}{\bar{l}_\text{c}}
\newcommand{\lexp}{l_\text{e}}
\newcommand{\rexp}{r_\text{e}}
\newcommand{\etaexp}{\eta_\text{e}}
\newcommand{\etab}{\eta_\text{b}}

\begin{document}

\title{Deterministic and statistical methods for the characterisation of poroelastic media from multi-observation sound absorption measurements}

\author{
  Jacques Cuenca$^{1}$, Peter Göransson$^2$, Laurent De Ryck$^1$, Timo Lähivaara$^3$\\\\
  \normalsize $^1$ Siemens Industry Software, Interleuvenlaan 68, BE-3001 Leuven, Belgium\\
  \normalsize $^2$ Department of Aeronautical and Vehicle Engineering,  KTH Royal Institute of Technology,\\
  \normalsize Teknikringen 8, SE-10044 Stockholm, Sweden\\
  \normalsize  $^3$ Department of Applied Physics, University of Eastern Finland, \\
  \normalsize  P.O. Box 1627, FIN-70211 Kuopio, Finland
}

\maketitle

\subsection*{Abstract}

This paper proposes a framework for the estimation of the transport
and elastic properties of open-cell poroelastic media based on sound
absorption measurements.  The sought properties are the
Biot-Johnson-Champoux-Allard model parameters, namely five transport
parameters, two elastic properties and the mass density, as well as
the sample thickness.  The methodology relies on a multi-observation
approach, consisting in combining multiple independent measurements
into a single dataset, with the aim of over-determining the problem.
In the present work, a poroelastic sample is placed in an impedance
tube and tested in two loading conditions, namely in a rigid-backing
configuration and coupled to a resonant expansion chamber.  Given the
non-monotonic nature of the experimental data, an incremental
parameter estimation procedure is used in order to guide the model
parameters towards the global solution without terminating at local
minima.  A statistical inversion approach is also discussed, providing
refined point estimates, uncertainty ranges and parameter
correlations.  The methodology is applied to the characterisation of a
sample of melamine foam and provides estimates of all nine parameters
with compact uncertainty ranges.  It is shown that the model
parameters are retrieved with a lower uncertainty in the
multi-observation case, as compared with a single-observation case.
The method proposed here does not require prior knowledge of the
thickness or any of the properties of the sample, and can be carried
out with a standard two-microphone impedance tube.

\section{Introduction}\label{sec:intro}
The estimation of material and geometrical properties of mechanical
and acoustical systems is central to many areas of engineering, yet
numerous methodological challenges exist.  One such example, of
particular interest here, is the measurement of transport and elastic
properties of poroelastic media.  These materials exhibit intricate
and complex fluid-structure interactions related to visco-thermal
exchanges and visco-elastic energy dissipation, and as a result are
often modelled with a large number of parameters.  The estimation of a
complete set of these parameters, which consistently and coherently
satisfy a comprehensive constitutive model, has been a topic of
increasing interest, for instance for its use in simulation tools, as
computational power has grown over the last decades.

\subsection{Direct methods for individual parameters}
A variety of methods for the characterisation of poroelastic media
have been developed, including direct, indirect and inverse methods.
Direct methods involve dedicated experimental rigs providing
individual estimations of a subset of properties, e.g.~porosity,
tortuosity, thermal and flow permeability, viscous and thermal
characteristic lengths, Young's modulus and mechanical loss
factor~\cite{ISO10534-2,ISO9053-1,ISO9053-2,LangloisPannetonAtalla2001JASA,Jaouen2008AA,LeclaireKeldersLauriksMelonBrownCastagnede1996lambdaAirHe,LeclaireKeldersLauriksGlorieuxThoen1996,LeclaireUmnovaHoroshenkovMaillet2003porosity,moussatovayraultcastagnede2001}.
These approaches are reliable and powerful in that a potentially high
degree of accuracy is reached thanks to an appropriate control of
experimental conditions and isolation of sources of error.
Nevertheless, using individually measured parameters in a
comprehensive simulation model may lead to inconsistencies.  For
instance, separate samples are often required for the different
dedicated setups, thereby introducing variability due to inhomogeneity
across samples.  Also, the availability and maintenance of a full
suite of test rigs is often impractical.  As an alternative, indirect
and inverse methods have been proposed, of which a selection is
recalled below.  An exhaustive review of all available methods is out
of scope here and the reader is referred to the recent review by
Horoshenkov~\cite{Horoshenkov2017review} and references therein for a
complementary overview.

\subsection{Indirect methods}
Indirect analytical methods rely on derivations of the constitutive
properties from the intrinsic equivalent acoustical properties of the
porous sample, namely its frequency-dependent equivalent density and
bulk modulus, which can be measured using a three- or four-microphone
transmission impedance tube, or a dual-load two-microphone tube.
First proposed for the estimation of viscous and thermal dissipation
properties by Panneton and
Olny~\cite{PannetonOlny2006,OlnyPanneton2008}, the methodology has
been reported by Bonfiglio and Pompoli~\cite{BonfiglioPompoli2013} to
provide all 5 parameters in the Johnson-Champoux-Allard (JCA)
model~\cite{JohnsonKoplikDashen1987,ChampouxAllard1991} and recently
by Jaouen et al.~\cite{JaouenGourdonGle2020estimation6paramsJCALtube}
for the measurement of all 6 parameters of the
Johnson-Champoux-Allard-Lafarge (JCAL)
model~\cite{JohnsonKoplikDashen1987,ChampouxAllard1991,LafargeLemarinierAllardTarnow1997JASA}.
Such an indirect methodology is robust and mature, and allows for a
comprehensive estimation of the full set of transport properties and
their uncertainties.  Nevertheless, these estimations are inherently
dependent on a number of assumptions as well as on the accuracy of the
measured intrinsic acoustical properties~\cite{PannetonOlny2006}.
Additionally, a minor drawback is that the estimation of the porosity
and flow resistivity rely on the manual identification of frequency
ranges where the intrinsic acoustical quantities exhibit an asymptotic
behaviour.  Although this could be automated, practical considerations
such as frequency resolution could influence the frequency ranges and
potentially lead to an underestimation of parameter uncertainties.
Complementary analytical approaches have been proposed, for instance
by Groby et al.~\cite{GrobyOgamDeryckSebaaLauriks2010} for estimating
transport properties from ultrasonic reflection and transmission
measurements.  The ultrasonic range inherently provides a direct
comparison with the high-frequency approximation of the JCA model but
does not provide access to the flow resistivity, thus requiring it to
be measured separately.

\subsection{Deterministic inverse methods}
Inverse methods stand as an appealing alternative to direct and
indirect methods in that the material properties are obtained as the
set of model parameters yielding the best fit of a simulation based on
the constitutive model.  This guarantees that the solution is a
feasible set of parameters that automatically satisfies the model,
without the need for approximations or additional steps.  A vast
number of deterministic inverse methods have been developed and
validated.  For instance, methods have been proposed to estimate all,
or a subset of, the transport parameters in the JCA or JCAL models
using the measured complex characteristic impedance as the target
data~\cite{AtallaPanneton2005inverse,BonfiglioPompoli2013,Zielinski2015inverse,GrobyDazelDeRyckKhanHoroshenkov2013graded},
which can be carried out in an impedance tube with transmission
capabilities.  Methods using a two-microphone tube have also been
investigated.  For instance, Dossi et
al.~\cite{DossiBrennanMoesenVandenbroeckLisha2019inverse} have
proposed the additional estimation of the elastic parameters of
polyurethane foams by combining datasets from specimens of different
thicknesses.  A method for estimating the 5 JCA parameters and 3
elastic properties has been proposed by Verdière et
al.~\cite{VerdierePannetonAtallaElkoun2017inverse} relying on a
numerical model accounting for intentional circumferential air gaps
and using sound absorption measurements, assuming prior knowledge of
the sample dimensions and density.  A numerical model was also used by
Vanhuyse et
al.~\cite{VanhuyseDeckersJonckheerePluymersDesmet2016inverse} to
account for intentionally clamped circumferential boundaries.  The
retrieval of transport and elastic parameters from the reflection
coefficient was numerically validated therein.  In addition to
impedance tube methods, deterministic inverse methods for the
estimation of anisotropic transport and elastic
properties~\cite{VanDerKelenGoransson2014JASA,VanDerKelenCuencaGoransson2015PT,CuencaVanderkelenGoransson2014JAP}
as well as methods relying on ultrasonic
measurements~\cite{FellahMitriFellahOgamDepollier2007inverse,DeryckLauriksLeclaireGrobyWirginDepollier2008reconstruction}
have been studied.  In particular, Ogam et
al.~\cite{OgamFellahSebaaGroby2011jsv} proposed an ultrasonic
time-domain method for the estimation of porosity, tortuosity, viscous
characteristic length, density, Young's modulus and Poisson ratio.

\subsection{Statistical inverse methods}
Fully deterministic approaches lack however the ability to estimate
uncertainty ranges for the model parameters.  Chazot et
al.~\cite{ChazotZhangAntoni2012jasa} proposed a Bayesian framework for
the estimation of 8 properties of poroelastic media, including mass
density, 5 transport parameters and 2 elastic properties, including
estimated confidence intervals for all the parameters.  As such, the
obtained solutions were not unique, and the mean estimates presented
large deviations from reference values in general, the reported
elasticity showing the largest estimation errors in particular.
Nevertheless, the method demonstrates the feasibility of fully
characterising poroelastic media using an impedance tube.  Niskanen et
al.~\cite{NiskanenGrobyDuclosDazelLerouxPoulainHuttunenLahivaara2017}
studied deterministic and statistical inverse methods for the
estimation of the 6 parameters of the JCAL model.  The method uses a
transmission tube setup and provides point estimates, credible ranges
and correlations between parameters using the equivalent density and
bulk modulus as the target data.  A two-microphone tube with a
rigid-backed sample was investigated as an experimentally simpler
alternative, where the difficulty of avoiding local minima was
reported, as well as a higher degree of inter-parameter correlation in
the obtained solution.  In further work, Niskanen et
al.~\cite{NiskanenDazelGrobyDuclosLahivaara2019jsv} discussed the
estimation of transport and elastic properties of poroelastic media
using reflection and transmission coefficients in the ultrasonic
range.  The work numerically demonstrates the feasibility of such an
approach and allows for the estimation of shear properties by exciting
the material in oblique incidence.  Other notable studies include the
works of Roncen et
al.~\cite{RoncenFellahSimonPiotFellahOgamDepollier2018bayesian,RoncenFellahLafargePiotSimonOgamFellahDepollier2018bayesian,RoncenFellahPiotSimonOgamFellahDepollier2019bayesviscous},
who investigated the statistical inference of the transport properties
of rigid porous media in the time domain, including a high order model
of viscous effects.  Fackler et
al.~\cite{FacklerXiangHoroshenkov2018bayesianmultilayer} used a
Bayesian framework for the characterisation of multi-layer rigid
porous media in terms of the porosity, flow resistivity, tortuosity
and thickness of the layers.

\subsection{Motivation for the proposed method}
The use of round robin tests has shown that acoustic measurements on
porous media exhibit relatively poor inter-laboratory
reproducibility~\cite{horoshenkov2007roundrobin,Pompoli2017roundrobin}.
The reported measurements included frequency-dependent characteristic
impedance, complex wavenumber and sound absorption coefficient.
Nevertheless, up to 3 times better reproducibility was reported for
sound absorption measurements in comparison to surface
impedance~\cite{horoshenkov2007roundrobin}, suggesting that the use of
sound absorption in combination with an inverse method can potentially
reduce uncertainties in the estimated parameters.  A closed-form
description of the behaviour of a rigidly backed isotropic poroelastic
layer exists~\cite{AllardAtalla2009}.  Assuming that this model
applies as well to samples with longitudinally-sliding boundaries in a
rigidly backed tube, it forms the basis for the model inversion in the
present work.  In this respect, the reduced amount of data inherent to
a purely energetic quantity, as compared to complex quantities, could
possibly add to the above-mentioned problems with parametric
insensitivity and the associated convergence of the solution to a
unique set of parameters.  This is particularly true in the case of
coupled resonant systems, whose inverse problems are subject to
multiple local minima due to the non-monotonic nature of their
frequency response.

Recent work by the authors proposes an incremental method for solving
inverse problems in coupled resonant systems from the knowledge of
their transfer function~\cite{GoranssonCuencaLahivaara2018mssp}.  The
method consists in solving the inverse problem in a series of steps,
starting with a sub-problem in a frequency domain where the observed
system exhibits an asymptotic behaviour, then gradually increasing the
frequency span towards that of the full problem.  Such an incremental
method was shown to guide the design variables towards the global
solution, while avoiding local minima.  This approach has also been
applied in preliminary numerical investigations aiming to estimate the
properties of porous and poroelastic
media~\cite{CuencaGoranssonDeryckLahivaara2018ISMA,CuencaDeryckGoranssonLahivaara2019ICSV},
laying the ground for the present work.

\subsection{Outline of the paper}
This paper proposes an inverse method for the simultaneous estimation
of a set of 9 properties of a poroelastic material sample using a
two-microphone impedance tube.  The target unknowns are the 5
transport parameters and 2 elastic properties within the
Biot-Johnson-Champoux-Allard (Biot-JCA)
model~\cite{AllardAtalla2009,JohnsonKoplikDashen1987,ChampouxAllard1991},
the material density and the sample thickness.  Here the previously
mentioned incremental estimation method is augmented with a
multi-observation approach, which is introduced with the aim of
over-determining the problem.  The multi-observation dataset consists
of measurements of the sound absorption coefficient of a rigidly
backed sample in two different loading conditions, namely with and
without an expansion chamber.  In order to ensure a gradual increase
of the problem complexity, the sequence of sub-problems is defined by
splitting the measured data into alternating monotonic
segments~\cite{Brooks1994}.  In the present paper, the multiple
sub-problems are solved in a deterministic sense using non-linear
numerical optimisation.  In order to assess the uncertainty of the
parameter estimation, a statistical method within the Bayesian
framework is then applied to the full problem as a means to obtain
credible ranges and refined point estimates for the different model
parameters.

The paper is organised as follows.  Section~\ref{sec:problem} presents
the proposed load cases, the model of the corresponding sound
absorption coefficient and the experimental setup used.
Section~\ref{sec:inversion} formalises the multi-observation model and
proposes a deterministic incremental inversion method as well as a
statistical inversion method.  The results of the approach are
presented in Sec.~\ref{sec:results} and a summary of the findings is
given in Sec.~\ref{sec:discussion}.

\section{Multi-observation problem}\label{sec:problem}

The aim of the work is to estimate the properties of a poroelastic
material (PEM) sample from the knowledge of its sound absorption
coefficient in one or various loading conditions.  In the following
paragraphs we propose a model of the sound absorption coefficient in
the different loading conditions as well as an experimental
implementation.

\subsection{Model}\label{sec:model}
The starting point is the model of the sound absorption coefficient in
the different load cases as a function of the unknown material
parameters of interest.  As the objective is to estimate these
parameters based on experimental data, the dimensions of the loading
elements, which are subject to uncertainties, are considered unknown
as well.  Furthermore, visco-thermal losses at the tube walls are
modelled using the approach in Ref.~\cite{bruneau2013}.

\begin{figure}[ht!]\centering
  \subfloat[]{\includegraphics{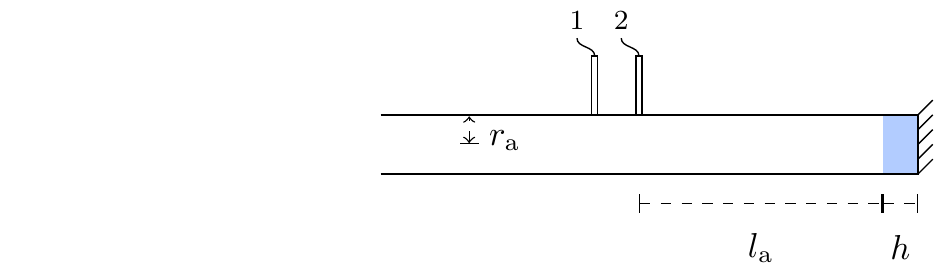}}\\
  \subfloat[]{\includegraphics{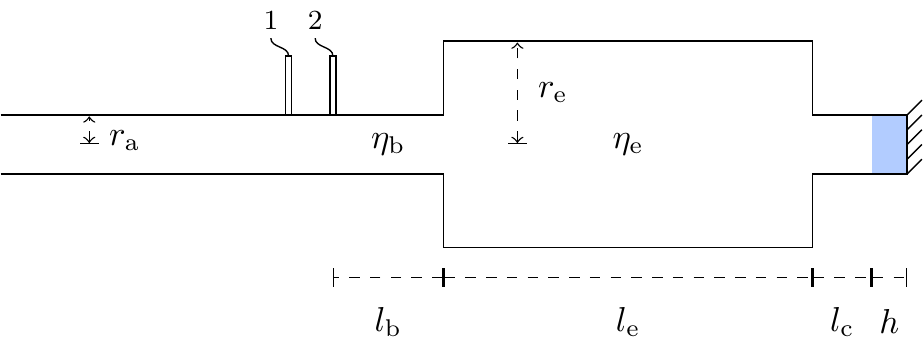}}
  \caption{Schematic representation of the impedance tube setup in the
    two loading conditions. (a) Load case \loadA: rigid-backed
    material sample; (b) load case \loadB: rigid-backed material
    sample coupled to expansion chamber.}
  \label{fig:model}
\end{figure}

The two loading conditions considered in this work are depicted in
Fig.~\ref{fig:model}.  Case \loadA\ consists of a rigidly-backed
poroelastic sample at the end of a tube of length $\la$.  In case
\loadB\, the sample is loaded with three adjacent cylindrical duct
elements of lengths $\lb$, $\lexp$ and $\lc$ forming an expansion
chamber of radius $\rexp$, as illustrated in the figure.  The tube of
length $\lc$ between the expansion chamber and the foam acts as a
coupling element that can be tuned so as to maximise the influence of
the elasticity of the foam on the overall sound absorption.

The coupled system is modelled as a multi-layer system using the
transfer matrix method~\cite{AllardAtalla2009}.  The rigid-backed
poroelastic material sample is modelled using the normal incidence
impedance derived in~\cite{AllardAtalla2009,Dazel2013HDR}, here
denoted
\begin{equation}
  \label{eq:Zpem}
  Z_\text{PEM}(\mathbf{x}_\text{PEM},\omega),
\end{equation}
where $\omega$ is the circular frequency and
\begin{equation}
  \label{eq:xpem}
  \mathbf{x}_\text{PEM} = \left\{ h,\rho,\phi,\sigma,\alpha_\infty,\Lambda,\Lambda',E,\eta \right\}
\end{equation}
is the set of parameters governing the behaviour of the poroelastic
sample, with $h$ the thickness, $\rho$ the mass density, $\phi$ the
porosity, $\sigma$ the static flow resistivity, $\alpha_\infty$ the
high-frequency limit of the tortuosity, $\Lambda$ the viscous
characteristic length, $\Lambda'$ the thermal characteristic length,
$E$ the Young's modulus and $\eta$ the viscoelastic damping ratio.
The estimation of these 9 parameters is the main objective of the
paper.  For the purposes of the present paper, the explicit expression
of Eq.~(\ref{eq:Zpem}) will not be repeated here, for which the reader
is referred to Refs.~\cite{AllardAtalla2009,Dazel2013HDR}.  It is
worth noting that due to the one-dimensional nature of the problem,
the Young's modulus and the Poisson ratio cannot be measured as two
independent variables.  Therefore lateral deformation is here ignored
by setting the Poisson ratio to zero.

For convenience in modelling the sudden area changes in load case
\loadB, the transfer matrix method is here formulated in terms of the
acoustic pressure and flow.  The transfer matrices of the individual
duct elements are given by~\cite{Munjal1987}
\begin{equation}
  \label{eq:Tn}
  \mathbf{T}_n =
  \begin{bmatrix} \cos(k_0l_n) &\dfrac{iZ_0}{\pi r_n^2}\sin(k_0l_n)\\ -\dfrac{\pi r_n^2}{iZ_0}\sin(k_0l_n)& \cos(k_0l_n) \end{bmatrix},
  \quad n=\text{a},\text{b},\text{e},\text{c},
\end{equation}
where $k_0=\omega/c_0$ is the wavenumber in air and $Z_0=\rho_0c_0$ is
the characteristic impedance of air, with $c_0$ the speed of sound and
$\rho_0$ the mass density of air.  The terms $l_n$ and $r_n$ denote
the length and radius of the various duct elements, with
$\ra=r_\text{b}=r_\text{c}$.  For load case \loadA, the transfer
matrix of the tube between microphone 2 and the sample is denoted
\begin{equation}
  \label{eq:T=Ta} \mathbf{T}^{\text{\loadA}}(\la) = \mathbf{T}_\text{a}.
\end{equation}
For load case \loadB, the joint transfer matrix of the tube section, the expansion chamber and the cavity is
\begin{equation}
  \label{eq:T=Tb.Te.Tc} \mathbf{T}^{\text{\loadB}}(\mathbf{x}_\text{load}) = \mathbf{T}_\text{b}\mathbf{T}_\text{e}\mathbf{T}_\text{c},
\end{equation}
where
\begin{equation}
  \label{eq:xload}
  \mathbf{x}_\text{load} = \left\{ \lb, \lexp, \rexp, \etab, \etaexp, \lc \right\}
\end{equation}
are the parameters governing the behaviour of the expansion chamber
and coupling tube.  The complete set of variables of interest for load
cases \loadA\ and \loadB\ can be denoted
\begin{align}
  \label{eq:x}
  \mathbf{x}^{\loadA}&=\mathbf{x}_\text{PEM},\\
  \mathbf{x}^{\loadB}&=\left\{ \mathbf{x}_\text{PEM}, \mathbf{x}_\text{load} \right\},
\end{align}
which consist of 9 and 15 unknown parameters, respectively.
The impedance of the coupled system is given by
\begin{equation}
  \label{eq:Z}
  Z_n(\mathbf{x}^{(n)},\omega) = \dfrac{ Z_\text{PEM}T^{(n)}_{11} + T^{(n)}_{12}S }{ Z_\text{PEM}T^{(n)}_{21}/S + T^{(n)}_{22} },\quad n=\text{A},\text{B},
\end{equation}
where $S = \pi\ra^2$ is the cross-section of the impedance tube and
$T^{(n)}_{ij}$ are the components of the transfer matrix of load
\loadA\ or \loadB.  The reflection coefficient of the compound system
is then
\begin{equation}
  \label{eq:R}
  R_n(\mathbf{x}^{(n)},\omega) = \dfrac{Z_n(\mathbf{x}^{(n)},\omega)-Z_0}{Z_n(\mathbf{x}^{(n)},\omega)+Z_0},
\end{equation}
yielding the absorption coefficient as
\begin{equation}
  \label{eq:alpha}
  \alpha_n(\mathbf{x}^{(n)},\omega) = 1-\left|R_n(\mathbf{x}^{(n)},\omega)\right|^2.
\end{equation}

\subsection{Experimental setup}\label{sec:experiment}

\begin{figure}[ht!]
  \centering
  \subfloat[]{\includegraphics[height=29mm]{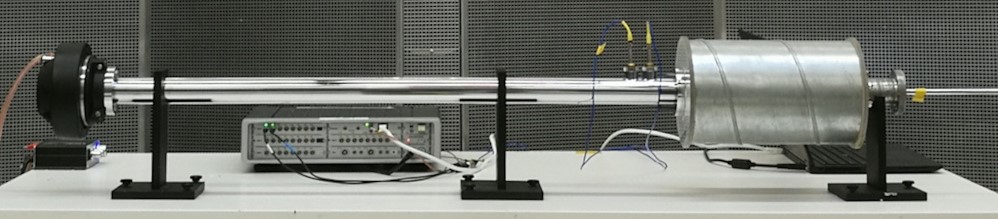}}\quad
  \subfloat[]{\includegraphics[height=29mm]{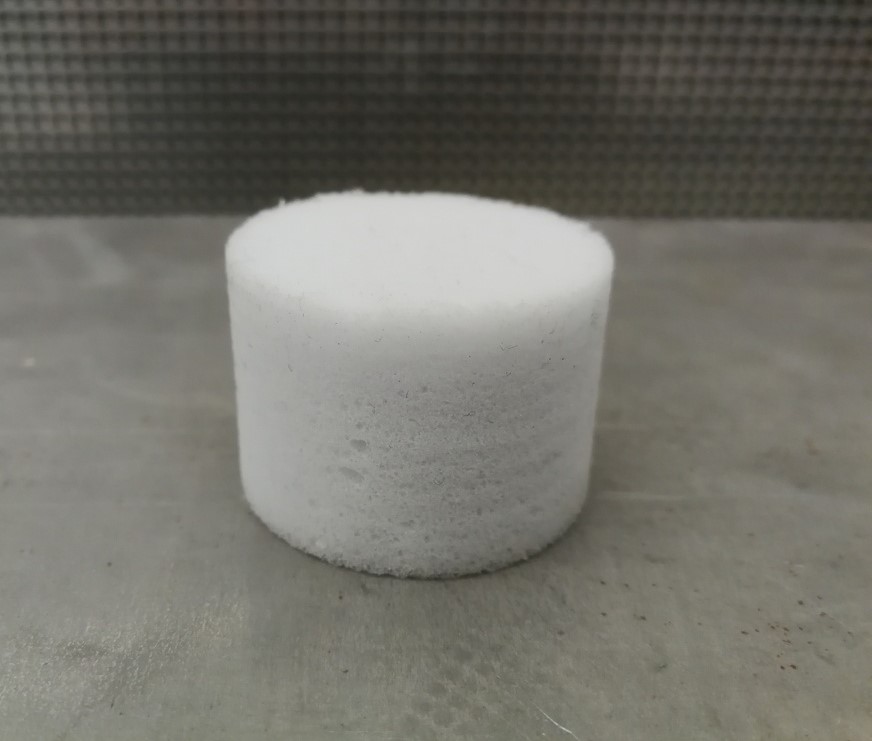}}
  \caption{Experimental setup and material sample. (a) Impedance tube with expansion chamber and poroelastic material termination; (b) melamine foam sample.}
  \label{fig:setup}
\end{figure}

This section describes the experimental setup and provides the nominal
values of the dimensions as depicted in Fig.~\ref{fig:model}, on which
the model depends.  Fig.~\ref{fig:setup} shows the impedance tube in
load case \loadB\ and the foam sample of interest.

The experimental setup relies on a Spectronics impedance tube with two
G.R.A.S.~46BD 1/4" microphones with nominal sensitivity $1.45\,$mV/Pa,
spaced by $29.21\,$mm, and the acquisition and processing are carried
out via a Simcenter SCADAS Mobile 05 in Simcenter Testlab 17 ``Sound
absorption testing using impedance tube'', following the procedure
described in the international standard~\cite{ISO10534-2}.  The
calibration transfer functions and all subsequent acquisitions are
performed under broadband random noise excitation using 100
non-overlapping Hann window averages.  The frequency resolution of the
measurements is $0.78125\,$Hz.

In order to achieve the loading condition illustrated in
Fig.~\ref{fig:model}(a), the sample is placed in a secondary tube
connected to the primary tube section.  The loading condition
illustrated in Fig.~\ref{fig:model}(b) is achieved by inserting an
expansion chamber between the primary and secondary tube sections.
The expansion chamber is made of aluminium, with acrylic lateral
walls.

The sample of melamine foam was cut by hand to a thickness of $h=24\pm
1\,$mm and attached to a rigid piston with double-sided tape.  The
piston comprises a seal around its circumference and is allowed to
slide along the tube.  The tube radius is provided by the manufacturer
as $\ra=17.43\,$mm and here assumed to be exact.  The distance from
microphone 2 to the modelled system was measured as $\la=80\pm5\,$mm
in load case \loadA\ and nominally as $\lbO=100\pm5\,$mm in load case
\loadB.  The inner dimensions of the expansion chamber were measured
as $\rexp=63\pm2\,$mm and $\lexp=227\pm2\,$mm.

The nominal length of the coupling tube between the expansion chamber
and the foam is $\lcO=40\pm1\,$mm.  Because of the sudden area changes
at the connection of this tube with the expansion chamber, a length
correction~\cite{Rayleigh,Ingard1953jasa,JaouenChevillotte2018aa} is
required in order to account for inertial effects.  The operational
length of the coupling tube is thus estimated as $\lc=51.7\pm3.6\,$mm,
where the lower and upper limits are estimated in a conservative
manner, corresponding to a tube of length $\lcO$ connected to an
expansion of radius $\rexp$ and to a baffled tube of length $\lcO+h$,
respectively.  A similar reasoning for the length between microphone 2
and the expansion chamber yields $\lb=111.7\pm7.6\,$mm.

Further required parameters are those related to the properties of
air, whose values are here considered exact, namely the speed of sound
$c_0=343\,$m$\cdot$s$^{-1}$, the mass density
$\rho_0=1.204\,$kg$\cdot$m$^{-3}$, the atmospheric pressure
$p_0=101320\,$Pa, the dynamic viscosity
$\mu_0=1.84\cdot10^{-5}\,$kg$\cdot$m$^{-1}\cdot$s$^{-1}$, the ratio of
specific heats $\gamma=1.4$, and Prandtl's number Pr$=0.71$.

\section{Model inversion method}\label{sec:inversion}

This section proposes an observation model and presents the different
inversion methods here developed.  In the present application,
individual observations consist of a measurement of the
frequency-dependent sound absorption coefficient in one of the two
loading conditions \loadA\ or \loadB, denoted $\alphameas_n$, with
$n=1$ (load case \loadA) or $n=2$ (load case \loadB).  In this
section, a generic notation $\gmeas_n$ with $n=1,\dots,N$ is preferred
for the sake of generality.

\subsection{Observation model}\label{sec:observation}

The experimental observation for a given load case $n$ at circular
frequency $\omega_m$ is denoted $\gmeas_n(\omega_m)$ and assumed
real-valued.  It is further assumed that observations for different
load cases are independent, that is, the outcome of any of the
observations does not affect the others.  It is then possible to
construct a compound measurement $\gmeasbold$ by concatenating
individual observations, as
\begin{equation}
  \label{eq:multi-observation}
  \gmeasbold =
  \begin{bmatrix}
    \begin{bmatrix}
      \gmeas_1(\omega_1) & \gmeas_1(\omega_2) & \cdots & \gmeas_1(\omega_M)
    \end{bmatrix}^\text{T}
    \\
    \begin{bmatrix}
      \gmeas_2(\omega_1) & \gmeas_2(\omega_2) & \cdots & \gmeas_2(\omega_M)
    \end{bmatrix}^\text{T}
    \\
    \vdots \\
    \begin{bmatrix}
      \gmeas_N(\omega_1) & \gmeas_N(\omega_2) & \cdots & \gmeas_N(\omega_M)
    \end{bmatrix}^\text{T}
  \end{bmatrix},
\end{equation}
where $M$ is the number of frequency lines and $N$ is the number of
observations, yielding a dataset of size $(MN\times1)$.  It is assumed
that the experimental observation differs from the model prediction by
an error, such that
\begin{equation}
  \label{eq:g0=g+eps}
  \gmeasbold = \mathbf{g}(\mathbf{x}_0) + \varepsilon,
\end{equation}
where $\mathbf{g}(\mathbf{x}_0)$ is the mathematical model, evaluated
at the true set of parameters $\mathbf{x}_0$, and $\varepsilon$ is a
random variable representing the modelling and measurement
uncertainties.  Here, we model $\varepsilon$ as Gaussian, with zero
mean and a diagonal covariance matrix
$\mathbf{\Gamma}_\varepsilon=\sigma_\varepsilon^2\mathbf{I}$, where
$\sigma_\varepsilon^2$ is the error variance and $\mathbf{I}$ denotes
the identity matrix of size $(MN\times MN)$.

Owing to the existence of the error $\varepsilon$, the true set of
model parameters $\mathbf{x}_0$ is inaccessible with complete
certainty.  The following paragraphs present different model inversion
paradigms aiming at obtaining an estimate $\widetilde{\mathbf{x}}_0$
of the model parameters $\mathbf{x}_0$.

\subsection{Deterministic inversion framework}\label{sec:deterministic}
A deterministic estimate of the model parameters can be obtained by
minimising the difference between the experimental observation and the
model prediction, as
\begin{equation}
  \label{eq:argmin(fobj)}
  \xOdet = \arg\min_\mathbf{x} \fobj(\mathbf{x}),
\end{equation}
where
\begin{equation}
  \label{eq:fobj}
  \fobj(\mathbf{x}) = \dfrac{1}{MN} \sum_{n=1}^N\sum_{m=1}^M \left( \gmeas_n(\omega_m) - g_n(\mathbf{x},\omega_m) \right)^2
\end{equation}
is the objective function, here defined as the residual sum of
squares.  The normalisation factor ${1}/{MN}$ is added in order to
enable comparing residuals $\fobj(\xOdet)$ for observations comprising
different numbers of frequency lines $M$ or load cases $N$.

The minimisation of the objective function may be performed with any
general purpose optimisation tool.  Two solvers are used in the
present work, namely a sequential quadratic programming algorithm
(SQP)~\cite{NocedalWright2006optimization} and the globally convergent
method of moving asymptotes (GCMMA)~\cite{Svanberg2002GCMMA}, as
detailed later-on in Sec.~\ref{sec:results}.  As a convergence and
stopping criterion, tolerance factors $\delta_{f_\text{obj}}$ and
$\delta_\mathbf{x}$ are set for the relative variation of the
objective function and variables over a number \nit\ of consecutive
iterations.

\subsection{Incremental inversion}\label{sec:incremental}
The above deterministic framework is here used within the incremental
method developed by the authors
in~\cite{GoranssonCuencaLahivaara2018mssp,CuencaGoranssonDeryckLahivaara2018ISMA,CuencaDeryckGoranssonLahivaara2019ICSV}.
The model inversion procedure therein relies on a gradual
complexification of the problem, starting from an asymptotic
observation and terminating at the full desired problem.  At each
stage of the approach, the corresponding sub-problem is solved and the
solution is used as the starting point for the next sub-problem.  For
frequency-dependent observations, the approach starts with a narrow
low-frequency observation whose upper frequency limit is incrementally
raised.

A common measure for the complexity of a real function is the number
of alternating strictly monotonic segments~\cite{Brooks1994}.  In the
present work, in order to ensure that the series of sub-problems yield
a gradual increase of the complexity of the inverse problem, the
frequency range is split into segments where the measured data is
monotonic.  The problem is solved from the lowest measured frequency
to the upper limit of each monotonic segment sequentially, until
reaching the full frequency range.  In the case of a multi-observation
dataset, the upper limits of all monotonic segments across the
different observations are used.

In the present work, the different steps in the incremental inversion
are performed using the deterministic inversion framework detailed in
Sec.~\ref{sec:deterministic}.  In addition, in order to capture the
low-frequency asymptotic behaviour of the system, an initial frequency
range is added as half the frequency span of the first monotonic
section.  Also, in order to ensure convergence of the estimated set of
parameters, a final sub-problem is appended, where the tolerance
factors are refined by a factor 10 and considered over a larger number
of iterations.

\subsection{Statistical inversion framework}\label{sec:mcmc}
In order to obtain a measure of the uncertainty on the model
parameters of interest, the unknown quantities are modelled as random
variables and the problem is formulated using a Bayesian
framework~\cite{KaipioSomersalo2005,calvettiSomersalo2007}.  The
latter allows to evaluate the set of unknowns in a probabilistic
sense, that is, by estimating their conditional probabilities for a
given observation $\gmeasbold$.  In this work, the unknown quantities
of interest are the model parameters $\mathbf{x}$.  In addition, the
standard deviation of the error, $\sigma_\varepsilon$, is considered
unknown, and modelled as a random variable as well.  The conditional
probability density of the set of unknowns for a given observation is
given by Bayes' formula~\cite{KaipioSomersalo2005},
\begin{equation}
  \label{eq:bayes}
  P(\mathbf{x},\sigma_\varepsilon|\gmeasbold) \sim P(\gmeasbold|\mathbf{x},\sigma_\varepsilon) P(\mathbf{x},\sigma_\varepsilon),
\end{equation}
where $P(\mathbf{x},\sigma_\varepsilon|\gmeasbold)$ is also referred
to as the posterior density and
$P(\gmeasbold|\mathbf{x},\sigma_\varepsilon)$ is the likelihood of the
observation given the unknowns.  The prior
$P(\mathbf{x},\sigma_\varepsilon)$ is here assumed uniform across a
sufficiently wide range within the admissible domain of the model
parameters and of the standard deviation of the error.  Owing to the
assumption of a Gaussian error $\varepsilon$ in the observation model
(\ref{eq:g0=g+eps}), the likelihood can be expressed as
\begin{eqnarray}
  P(\gmeasbold|\mathbf{x},\sigma_\varepsilon) &=& P_\varepsilon(\mathbf{g}(\mathbf{x})-\gmeasbold)\nonumber\\
  \label{eq:likelihood_general}
                                              &=& \frac{1}{\sqrt{\left(2\pi\right)^{MN}\det\left(\mathbf{\Gamma}_\varepsilon\right)}}
                                                  \exp\left(-\frac{1}{2}\left(\gmeasbold-\mathbf{g}(\mathbf{x})\right)^\text{T}\mathbf{\Gamma}_\varepsilon^{-1}\left(\gmeasbold-\mathbf{g}(\mathbf{x})\right)\right),
\end{eqnarray}
where $P_\varepsilon$ is the probability density of the error.
Furthermore, the assumption that the covariance matrix of the error is
diagonal yields
\begin{equation}
  \label{eq:likelihood_detailed}
  P(\gmeasbold|\mathbf{x},\sigma_\varepsilon)
  = \frac{1}{\left(\sqrt{2\pi}\sigma_\varepsilon\right)^{MN}}
  \exp\left(-\frac{1}{2\sigma_\varepsilon^2}\sum_{n=1}^N\sum_{m=1}^M \left( \gmeas_n(\omega_m) - g_n(\mathbf{x},\omega_m) \right)^2\right).
\end{equation}

In the present work, the posterior density is sampled by means of a
Markov Chain Monte Carlo method, using the Metropolis-Hastings
algorithm with an adaptive proposal
scheme~\cite{HaarioSaksmanTamminen1999,HaarioSaksmanTamminen2001}.
The purpose of such a procedure is to visit the admissible domain of
the unknowns with a probability
$P(\mathbf{x},\sigma_\varepsilon|\gmeasbold)$, where the jumps between
successive points of the chain are determined by an adaptive proposal
distribution.  In practice, $\sigma_\varepsilon$ is estimated together
with the model parameters $\mathbf{x}$, such that the complete set of
unknowns can be denoted
\begin{equation}
  \label{eq:y=[x,sigma_eps]}
  \mathbf{y}=\left\{\mathbf{x},\sigma_\varepsilon\right\}.
\end{equation}
The proposal distribution is here defined as Gaussian, where its
covariance matrix is initialised as diagonal and then adapted as a
function of the covariance matrix of past
samples~\cite{HaarioSaksmanTamminen2001}, as
\begin{equation}
  \label{eq:cov(x)}
  \mathbf{C}^{(j)}=\gamma_j\left(\text{cov}\left(\mathbf{y}^{(0)},\dots,\mathbf{y}^{(j-1)}\right)+\delta\mathbf{I}\right),\quad j\geq J,
\end{equation}
where $J$ is an arbitrary jump number at which the adaptation is
started, $\delta$ is an arbitrarily small number and $\mathbf{I}$
denotes the identity matrix of size $(D\times D)$, $D$ being the
dimension of the problem, i.e.~$D=10$ for the single-observation case
(load \loadA) and $D=16$ for the dual-observation case (loads
\loadA\ and \loadB).  The scaling parameter $\gamma_j$ is initialised
at~\cite{GelmanRobertsGilks1996MetropolisJumpingRules}
\begin{equation}
  \label{eq:gamma=2.38^2/D}
  \gamma_J=2.38^2/D
\end{equation}
and then updated following guidelines by Andrieu and
Thoms~\cite{AndrieuThoms2008tutorialMCMC} in order to reach the
desired acceptance rate $a_*$.  Here we choose
\begin{equation}
  \label{eq:gamma_{j+1}=fct(gamma_j)}
  \gamma_{j+1}=\gamma_j\mu_{j+1}^{a_j-a_*},\quad j\geq J,
\end{equation}
where $a_j$ is the acceptance rate at jump $j$ and $\mu_j$ is an
arbitrary strictly decreasing sequence converging to 1.

The solution provided by the deterministic inversion procedure,
$\xOdet$, is a feasible set of model parameters and therefore used as
a starting point for the chain.  This has the advantage of reducing
the required number of iterations in the burn-in phase.  Accordingly,
an initial estimate of the standard deviation $\sigma_\varepsilon$ of
the error is obtained from Eq.~(\ref{eq:g0=g+eps}) as
\begin{equation}
  \label{eq:var_eps}
  \widetilde\sigma_\varepsilon = \sqrt{\frac{1}{MN}\sum_{n=1}^N\sum_{m=1}^M\left( \gmeas_n(\omega_m) - g_n\left(\xOdet,\omega_m\right) \right)^2}.
\end{equation}
Alternatively, such an initial estimate can be obtained from the
spread of a series of repeated measurements, as suggested in
Ref.~\cite{NiskanenGrobyDuclosDazelLerouxPoulainHuttunenLahivaara2017}.
However, as this does not quantify modelling uncertainties it is not
suitable for the present study, and in particular for load case
\loadB, where the one-dimensional approximation of the expansion
chamber is known to have
limitations~\cite{Munjal1987,AlkmimCuencaDeryckGoransson2018ISMA}.

In the present work, the aim is to obtain point estimates and
uncertainty ranges for all unknowns.  We use the maximum a posteriori
estimate (MAP), which maximises the posterior probability density, the
conditional mean (CM) of the unknowns, and the narrowest 95\% credible
intervals~\cite{KaipioSomersalo2005}.

\section{Results}\label{sec:results}
This section presents the results of the parameter estimation for the
melamine foam sample of interest, introduced in
Sec.~\ref{sec:experiment}.  The target experimental data consists of a
single-observation dataset and a dual-observation dataset.  The
single-observation dataset consists of the sound absorption
coefficient of the foam in load case \loadA, and the dual-observation
dataset is composed of both the sound absorption coefficients
corresponding to load cases \loadA\ and \loadB, concatenated as
specified in Eq.~(\ref{eq:multi-observation}).

The following four paragraphs respectively present the incremental inversion results, a robustness study, the statistical inversion results and a comparison of multi-observation vs.\ single-observation estimations.

\subsection{Incremental model inversion}\label{sec:results_deterministic}
The full frequency range of interest for load case \loadA\ is here
defined as $f\in[200,4500]\,$Hz.  The low and high frequency limits
are respectively chosen so as to avoid uncertainties in the
inter-microphone phase and to limit the influence of the non-planar
modes of the tube, whose first transversal resonance corresponds to
the first azimuthal mode, i.e.~(1,0)~\cite{Munjal1987}, at circa
$5.7\,$kHz for the present setup.  For load case \loadB, the main tube
column imposes a non-zero transversal velocity along the central line
of the expansion chamber and therefore azimuthal modes are not
excited.  The second transversal resonance of the expansion chamber
occurs at the first radial mode, i.e.~(0,1), at circa $2.65\,$kHz
here, and therefore the upper frequency limit is chosen at $2\,$kHz.

Table~\ref{tab:frequency-intervals} shows the frequency ranges
obtained for the single observation and for the dual observation using
the procedure described in Sec.~\ref{sec:incremental}.
\begin{table}[ht!]
  \caption{Frequency intervals and tolerance criteria for the incremental inversion. (a) single observation; (b) dual observation.}
  \label{tab:frequency-intervals} 
  \centering
  \subfloat[]{
    \begin{tabular}[t]{cS[table-format=3.2]S[table-format=3.2]ccc}
      Step & $f_\text{min}$ & {\protect$f_\text{max}$ (Hz)} & $\delta_{f_\text{obj}}$ & $\delta_\mathbf{x}$ & \nit \\
      \midrule
      1   & 200 &   669.53 & $10^{-2}$ & $10^{-3}$ & 2 \\
      2   & 200 &  1139.06 & $10^{-2}$ & $10^{-3}$ & 2 \\
      3   & 200 &  1351.56 & $10^{-2}$ & $10^{-3}$ & 2 \\
      4   & 200 &  3964.06 & $10^{-2}$ & $10^{-3}$ & 2 \\
      5   & 200 &  4500 & $10^{-2}$ & $10^{-3}$ & 2 \\
      6   & 200 &  4500 & $10^{-3}$ & $10^{-4}$ & 4 \\
      \bottomrule
    \end{tabular}
  }
  \qquad
  \subfloat[]{
    \begin{tabular}[t]{cS[table-format=3.2]S[table-format=3.2]ccc}
      Step & {\protect$f_\text{min}$ (Hz)} & {\protect$f_\text{max}$ (Hz)} & $\delta_{f_\text{obj}}$ & $\delta_\mathbf{x}$ & \nit \\
      \midrule
      1 & 200 &   469.53 & $10^{-2}$ & $10^{-3}$ & 2 \\
      2 & 200 &   739.06 & $10^{-2}$ & $10^{-3}$ & 2 \\
      3 & 200 &   889.06 & $10^{-2}$ & $10^{-3}$ & 2 \\
      4 & 200 &  1114.06 & $10^{-2}$ & $10^{-3}$ & 2 \\
      5 & 200 &  1339.06 & $10^{-2}$ & $10^{-3}$ & 2 \\
      6 & 200 &  1526.56 & $10^{-2}$ & $10^{-3}$ & 2 \\
      7 & 200 &  1901.56 & $10^{-2}$ & $10^{-3}$ & 2 \\
      8 & 200 &  3964.06 & $10^{-2}$ & $10^{-3}$ & 2 \\
      9 & 200 &  4500 & $10^{-2}$ & $10^{-3}$ & 2 \\
      10 & 200 &  4500 & $10^{-3}$ & $10^{-4}$ & 4 \\
      \bottomrule
    \end{tabular}
  }
\end{table}

\begin{figure}[ht!]
  \centering
  \subfloat[]{\igh{51mm}{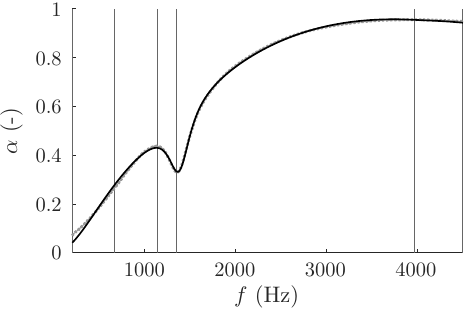}}\quad
  \subfloat[]{\igh{51mm}{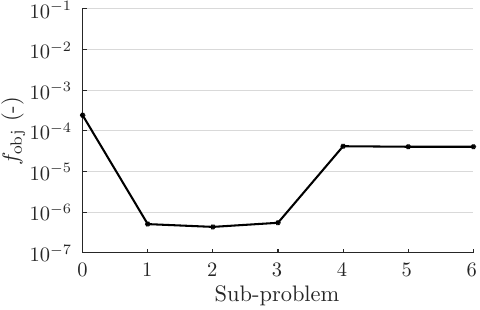}}
  \caption{Deterministic model inversion solution for a
    single-observation measurement of a rigid-backed poroelastic
    material in loading condition \loadA. (a) Sound absorption
    coefficient, {\protect\greydottedline} measurement,
    {\protect\blackline} model; (b) objective function history. The
    vertical lines represent the upper limits of the successive model
    inversion steps.}
  \label{fig:gcmma_single}
\end{figure}

\begin{figure}[ht!]
  \centering
  \subfloat[]{\igh{51mm}{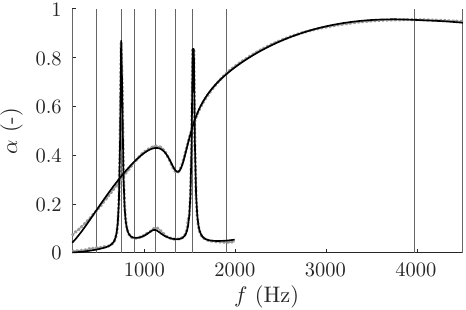}}\quad
  \subfloat[]{\igh{51mm}{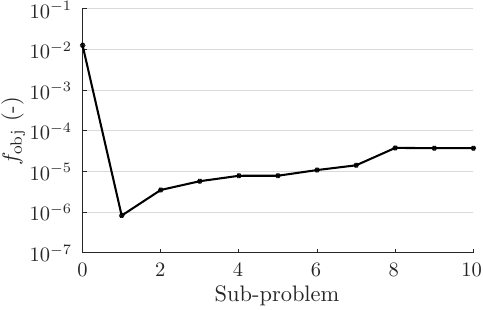}}
  \caption{Deterministic model inversion solution for the
    dual-observation measurement, including poroelastic material in
    loading condition \loadA\ and poroelastic material coupled to
    expansion chamber in loading condition \loadB. (a) Sound
    absorption coefficient, {\protect\greydottedline} measurement,
    {\protect\blackline} model; (b) objective function history. The
    vertical lines represent the upper limits of the successive model
    inversion steps.}
  \label{fig:gcmma_dual}
\end{figure}

Figures~\ref{fig:gcmma_single} and~\ref{fig:gcmma_dual} show the
measured and estimated sound absorption coefficient for the single-
and dual-observation cases, respectively, using a GCMMA implementation
of the proposed incremental inversion approach with a random starting
set of parameters.  A good approximation of the measurement is
achieved with the proposed model in both cases, with minor
discrepancies at the low-frequency limit.  These may be attributed to
the model's inability to capture low-frequency dissipation effects in
the impedance tube, or to the known limitations of the porous material
model here used~\cite{LafargeLemarinierAllardTarnow1997JASA}.  The
figures depict as well the objective function as a function of the
incremental inversion steps, and show generally increasing values due
to the increasing sub-problem complexity.

The figures clearly show the effect of the elasticity of the frame on
the sound absorption coefficient, observed for load case \loadA\ as a
dip at the elastic resonance of the material at $1351.56\,$Hz.  The
elastic effects also play an important role in load case \loadB, and
arise as a peak at $1114.06\,$Hz.  This is interpreted as a coupling
effect between the expansion chamber and the material frame via the
coupling tube.  It is worth noting that a sufficient degree of
sensitivity of load case \loadB\ to the properties of the foam is
crucial for the over-determination of the problem, as will be further
explored in Sec.~\ref{sec:estimates}.

\subsection{Robustness study}\label{sec:robustness}
\begin{figure}[ht!]
  \centering
  \subfloat[]{\ig{.83}{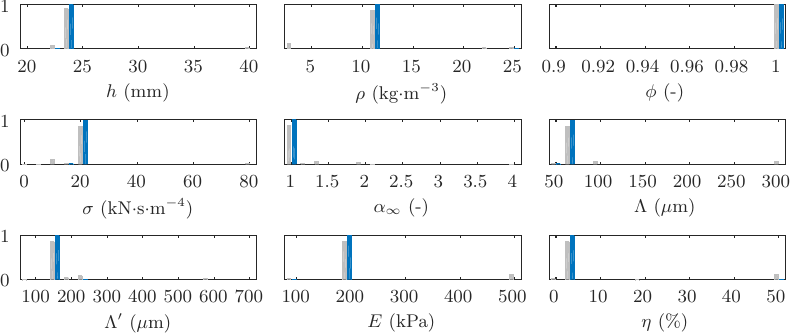}}\\
  \subfloat[]{\igh{.1079\linewidth}{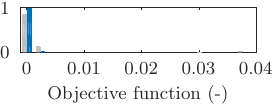}}\quad
  \subfloat[]{\igh{.1079\linewidth}{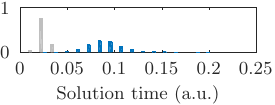}}
  \caption{Normalised histograms of parameter values, residual values
    and solution time for 1000 runs of the gradient-based
    single-observation case. Light grey, Full-range search; blue,
    incremental search.}
  \label{fig:hist_single_observation}
\end{figure}

\begin{figure}[ht!]
  \centering
  \subfloat[]{\ig{.83}{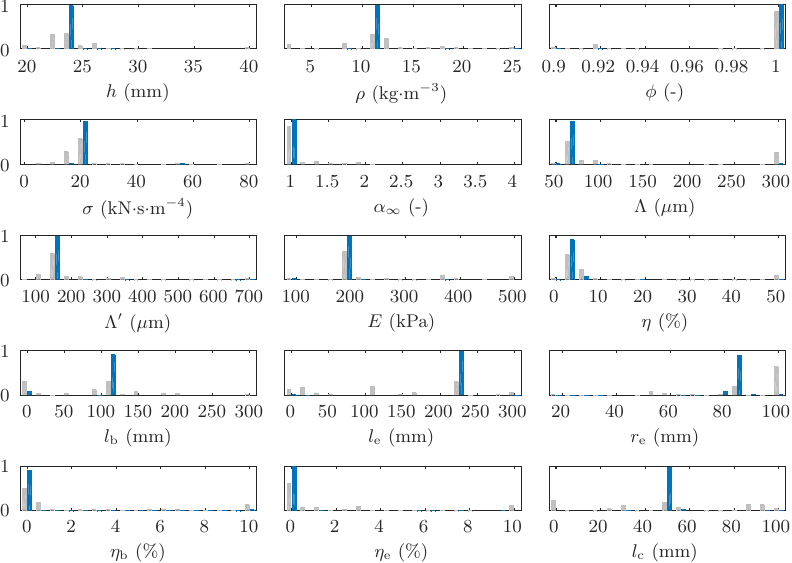}}\\
  \subfloat[]{\igh{.1079\linewidth}{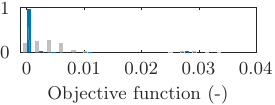}}\quad
  \subfloat[]{\igh{.1079\linewidth}{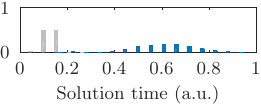}}
  \caption{Normalised histograms of parameter values, residual values
    and solution time for 1000 runs of the gradient-based
    multi-observation case. Light grey, Full-range search; blue,
    incremental search.}
  \label{fig:hist_dual_observation}
\end{figure}

A multi-start approach~\cite{Glover2006HandbookMetaheuristics} is here
used in order to evaluate the robustness of the incremental search
method, regardless of the initial guess of the parameter values.  The
procedure consists in repeating the model inversion process a number
of times, here using a uniform random distribution for the parameter
values as the starting point.
Figures~\ref{fig:hist_single_observation}
and~\ref{fig:hist_dual_observation} show the distribution of the
parameter values obtained for 1000 runs of a full-range search versus
the incremental search, for the single- and dual-observation cases
respectively, and presented in the form of histograms.  In addition,
the distribution of objective function residuals and the solution time
are also represented as histograms.  An implementation based on SQP
was preferred for this iterative procedure as the algorithm involves a
lower computational overhead than the authors' implementation of
GCMMA.

It can be observed that the incremental parameter search performs more
consistently than the full-range search, especially in the
dual-observation configuration.  The appearance of multiple solutions
elucidates the existence of local minima, which in turn results in a
widening of the distribution of the objective function residuals.
Nevertheless, the proposed incremental approach comes at a
non-negligible computational cost.  In the present case the expected
solution time is approximately fivefold with respect to the full-range
search.

\subsection{Statistical model inversion}\label{sec:results_bayes}
This paragraph presents the results of the Bayesian inversion
approach, consisting in sampling the posterior density, see
Sec.~\ref{sec:mcmc}.  For the present work, the target acceptance rate
is set to $a_*=0.2$.  The Markov Chain Monte Carlo sampling algorithm
is here initialised at the solution obtained from the incremental
inversion.  Although such a starting point is a feasible solution, a
burn-in phase is considered in order to gather a sufficient number of
samples for a reliable initial estimate of the covariance matrix of
the unknowns.  After the burn-in phase, the total number of samples
kept for the analysis is set to $2\cdot10^6$.  For the purposes of the
problem of interest, two point estimates are extracted, namely the
maximum a posteriori (MAP) and the conditional mean (CM).  In
addition, the 95\% credible ranges of each model parameter are
extracted.

\begin{figure}[ht!]
  \centering
  \ig{1}{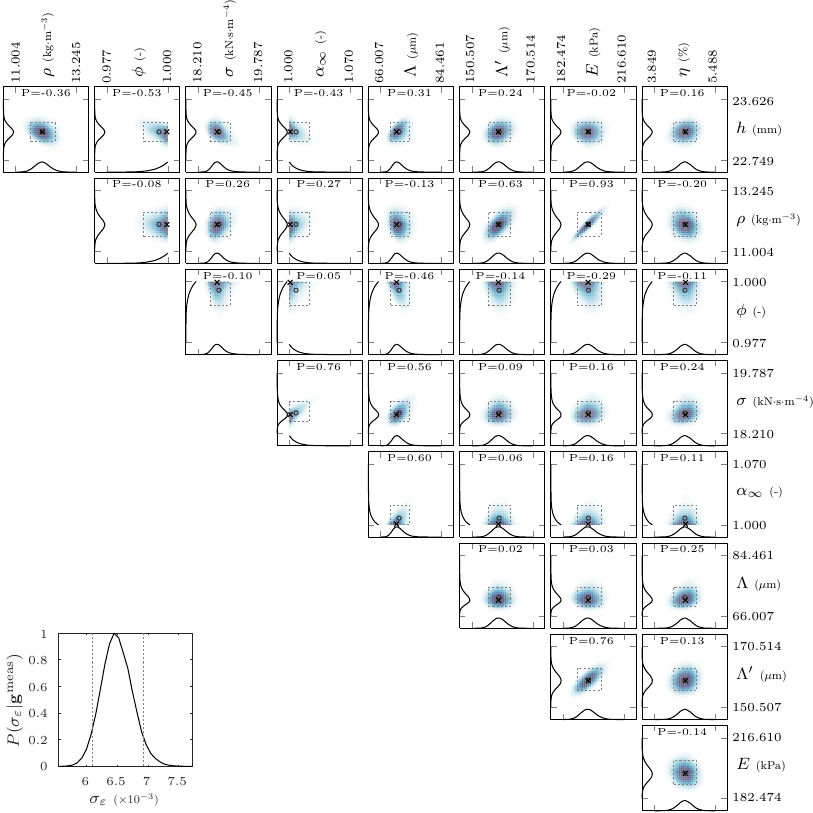}
  \caption{Pairwise posterior marginal densities for the
    single-observation case. {\protect\blackline}~individual marginal
    densities; {\protect\credline}~credible ranges;
    {\protect\MAPcross}~MAP estimate; {\protect\CMcircle}~CM estimate;
    $\Rho$,~Pearson correlation coefficient. Bottom left: Posterior
    marginal density of the error $\varepsilon$.}
  \label{fig:mcmc_single}
\end{figure}

Figure~\ref{fig:mcmc_single} shows the pairwise 2D marginal densities
for the 9 parameters of the poroelastic sample.  The figure also
includes the limits of the 95\% credible ranges as well as the 1D
marginal densities of the individual parameters.  In addition, a
measure of the inter-parameter correlation is provided thanks to the
Pearson correlation coefficient, represented with the symbol $\Rho$.
It can be observed that the support of the densities is compact and
presents a unique maximum point.  Indeed, thanks to the incremental
inversion solution being used as a feasible starting point, the
mapping of other regions of the parameter space is avoided, and
accordingly no multi-modal behaviour is observed.

A number of notable correlations among the parameters can be observed.
For instance, the sample thickness exhibits a higher correlation with
the acoustical transport properties than with the elastic properties.
Indeed, the elasticity of the frame does not contribute to a global
increase or decrease of the sound absorption coefficient, but locally
via the frame resonance.  Furthermore, the Young's modulus $E$ and the
mass density $\rho$ are positively correlated, which is expected due
to their joint control of the elastic resonance.  The flow resistivity
$\sigma$, the tortuosity $\alpha_\infty$ and the viscous
characteristic length $\Lambda$ are strongly correlated, as small
variations in their values contribute to sound absorption in a
comparable manner.  In fact, a similar trend in the correlation of
$\alpha_\infty$ and $\Lambda$ has been reported
\cite{ChazotZhangAntoni2012jasa}.  In addition to these correlations,
the thermal characteristic length $\Lambda'$ and the Young's modulus
$E$ exhibit a strong correlation.  Indeed, for the melamine foam here
tested, the thermal and elastic effects are both dominant over a
common mid-frequency range.

These correlations show the existence of a local interdependence of
the model parameters.  Several authors have reported on the scaling
laws governing the constitutive properties of poroelastic media over
small variations of the
parameters~\cite{GibsonAshby1999,Goransson2006,LindnordgrenGoransson2010,CameronLindnordgrenWennhageGoransson2014}.
For instance these scaling laws predict a positive correlation between
the density and the mass density, between the flow resistivity and the
tortuosity, and between the tortuosity and the viscous characteristic
length, such as observed here.  In addition to being intrinsic to the
constitutive model, these parameter correlations may also be
exacerbated by the choice of the sound absorption coefficient as the
target data, as reported by Niskanen et
al.~\cite{NiskanenGrobyDuclosDazelLerouxPoulainHuttunenLahivaara2017}.

The correlations observed here suggest that the model parameters are
not independent of one another.  As a matter of fact, a set of 3
independent parameters has been shown by Horoshenkov et
al.~\cite{HoroshenkovHurrellGroby2019jasa} to be sufficient to
correctly represent the acoustical behaviour of porous media,
classically modelled using the 5-parameter JCA model
herein~\cite{JohnsonKoplikDashen1987,ChampouxAllard1991}, or the
6-parameter Johnson-Champoux-Allard-Lafarge (JCAL)
model~\cite{JohnsonKoplikDashen1987,ChampouxAllard1991,LafargeLemarinierAllardTarnow1997JASA}.

\begin{figure}[ht!]
  \centering
  \ig{1}{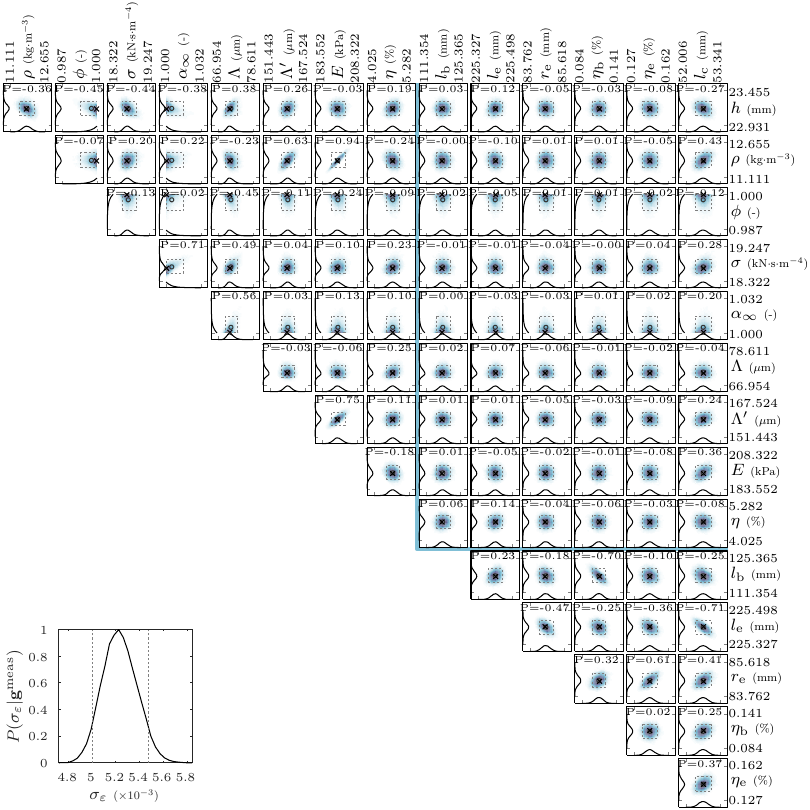}
  \caption{Pairwise posterior marginal densities for the
    dual-observation case. {\protect\blackline}~individual marginal
    densities; {\protect\credline}~credible ranges;
    {\protect\MAPcross}~MAP estimate; {\protect\CMcircle}~CM estimate;
    $\Rho$,~Pearson correlation coefficient;
    {\protect\thicktealline}~delimitation of pairwise densities
    pertaining to the sample only, the load only, and the coupling of
    the two. Bottom left: Posterior marginal density of the error
    $\varepsilon$.}
  \label{fig:mcmc_dual}
\end{figure}

Figure~\ref{fig:mcmc_dual} shows the pairwise marginal densities for
the dual-observation case.  The additional parameters exhibit compact
marginal densities and unique maxima as well, and introduce additional
correlations.  For instance, a high degree of correlation can be
observed between $\lb$ and $\etab$, which respectively account for
low-frequency and broadband dissipation in the impedance tube.
Similarly, $\rexp$ and $\etaexp$ respectively account for absorption
peak depth and broadband dissipation in the expansion chamber.
Overall, a low correlation between the marginal densities of the
poroelastic material parameters and those of the expansion chamber is
observed.  However, the coupling tube length $\lc$ exhibits a somewhat
higher correlation with $E$ and $\rho$.  This bears witness to the
fact that the coupling tube strongly interacts with the elastic
deformation of the foam.

\subsection{Multi-observation vs.\ single-observation estimations}\label{sec:estimates}
\begin{figure}[hbt!]
  \centering
  \ig{.92}{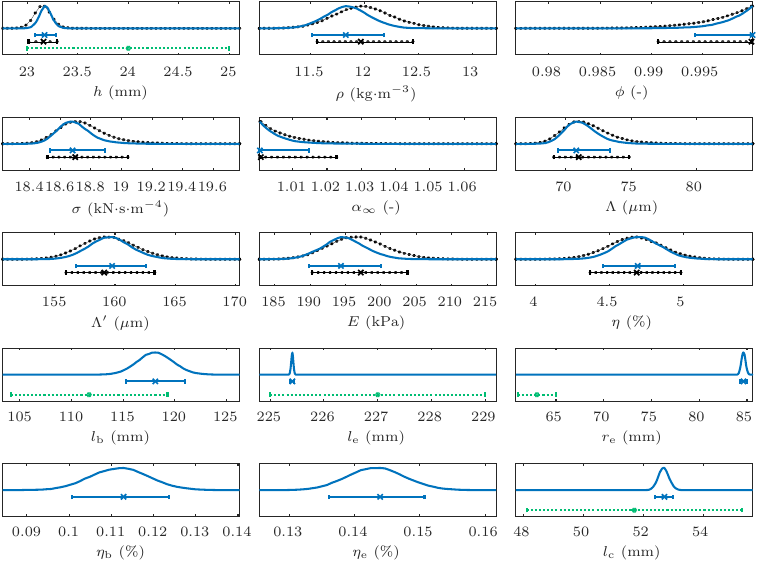}
  \caption{Individual marginal densities for all model
    parameters. {\protect\blackdottedline} Single-observation,
    {\protect\blueline} dual-observation. The horizontal error bars
    represent the credible ranges and the crosses denote the MAP
    estimate. {\protect\tealline} Manual length measurement.}
  \label{fig:mcmc_1D}
\end{figure}

Table~\ref{tab:solution} displays the parameter values obtained for
the deterministic incremental inversion as well as the MAP and CM
estimates and the 95\% credible intervals, for both the single- and
dual-observation cases.  In addition, Fig.~\ref{fig:mcmc_1D} shows the
marginal densities for all individual parameters, superimposed for the
single- and dual-observation cases, as well as the credible intervals
and MAP estimate values.
\begin{table}[hbt!]\centering
  \caption{Parameter values obtained from the deterministic
    incremental estimation and from the statistical inversion,
    including MAP estimate, CM estimate and 95\% credible intervals.}
  \label{tab:solution}
  \subfloat[]{
    \begin{tabular}{llS[table-format=3.4]S[table-format=3.4]S[table-format=3.4]S[table-format=3.4]S[table-format=3.4]}
      \textbf{Parameter} & \textbf{Unit} &\textbf{Determ.}&\textbf{MAP}&\textbf{CM}&\textbf{Cred. low}&\textbf{Cred. high}\\
      \midrule
      $h$                      & mm                            & 23.14       & 23.14       & 23.15       & 23.01       & 23.29       \\
      $\rho$                   & kg$\cdot$m$^{-3}$            & 11.865      & 11.958      & 11.994      & 11.562      & 12.449      \\
      $\phi$                   & -                             & 1.0000      & 0.9996      & 0.9966      & 0.9906      & 0.9998      \\
      $\sigma$                 & kN$\cdot$s$\cdot$m$^{-4}$   & 18.594      & 18.709      & 18.748      & 18.516      & 19.044      \\
      $\alpha_\infty$          & -                             & 0001        & 1.001       & 1.008       & 1           & 1.023       \\
      $\Lambda$                & $\mu$m                       & 69.53       & 70.93       & 71.61       & 69.09       & 74.82       \\
      $\Lambda'$               & $\mu$m                       & 158.66      & 159.07      & 159.46      & 155.92      & 163.24      \\
      $E$                      & kPa                           & 194.199     & 196.697     & 196.715     & 190.276     & 203.756     \\
      $\eta$                   & \%                          & 4.618       & 4.696       & 4.671       & 4.364       & 4.982       \\
      $\sigma_\varepsilon$     & $10^{-3}$                     & 6.364       & 6.447       & 6.492       & 6.099       & 6.916       \\
      \bottomrule
    \end{tabular}
  }\\
  \subfloat[]{
    \begin{tabular}{llS[table-format=3.4]S[table-format=3.4]S[table-format=3.4]S[table-format=3.4]S[table-format=3.4]}
      \textbf{Parameter} & \textbf{Unit} &\textbf{Determ.}&\textbf{MAP}&\textbf{CM}&\textbf{Cred. low}&\textbf{Cred. high}\\
      \midrule
      $h$                      & mm                            & 23.17       & 23.18       & 23.18       & 23.08       & 23.29       \\
      $\rho$                   & kg$\cdot$m$^{-3}$            & 11.766      & 11.821      & 11.846      & 11.512      & 12.190      \\
      $\phi$                   & -                             & 1.0000      & 0.9998      & 0.9980      & 0.9942      & 0.9999      \\
      $\sigma$                 & kN$\cdot$s$\cdot$m$^{-4}$   & 18.594      & 18.681      & 18.694      & 18.529      & 18.899      \\
      $\alpha_\infty$          & -                             & 0001        & 1.001       & 1.005       & 1           & 1.015       \\
      $\Lambda$                & $\mu$m                       & 69.91       & 70.65       & 71.22       & 69.38       & 73.47       \\
      $\Lambda'$               & $\mu$m                       & 159.07      & 159.75      & 159.62      & 156.75      & 162.62      \\
      $E$                      & kPa                           & 193.228     & 194.352     & 194.792     & 189.792     & 200.025     \\
      $\eta$                   & \%                          & 4.662       & 4.716       & 4.702       & 4.456       & 4.950       \\
      $l_\text{b}$             & mm                            & 118.07      & 118.13      & 118.13      & 115.33      & 121.00      \\
      $l_\text{e}$             & mm                            & 225.41      & 225.42      & 225.41      & 225.38      & 225.45      \\
      $r_\text{e}$             & mm                            & 84.62       & 84.60       & 84.63       & 84.27       & 84.99       \\
      $\eta_\text{b}$          & \%                          & 0.11        & 0.11        & 0.11        & 0.10        & 0.12        \\
      $\eta_\text{e}$          & \%                          & 0.143       & 0.143       & 0.143       & 0.136       & 0.151       \\
      $l_\text{c}$             & mm                            & 52.61       & 52.68       & 52.68       & 52.38       & 52.98       \\
      $\sigma_\varepsilon$     & $10^{-3}$                     & 6.091       & 5.224       & 5.235       & 5.003       & 5.477       \\
      \bottomrule
    \end{tabular}
  }
\end{table}

The obtained point estimates provide reasonable values for the
different parameters of the poroelastic material model.  Moreover,
these are consistent with those reported elsewhere for melamine
foams~\cite{NiskanenGrobyDuclosDazelLerouxPoulainHuttunenLahivaara2017,VanDerKelenGoransson2014JASA,CuencaVanderkelenGoransson2014JAP,VanDerKelenCuencaGoransson2015PT,Pompoli2017roundrobin,GeebelenBoeckxVermeirLauriksAllardDazel2007AAUA,Jaouen2008AA,BoeckxLeclaireKhuranaGlorieuxLauriksAllard2005investigation}.

The most notable result of the present work is that the credible
intervals obtained by the proposed method are narrower in the
dual-observation case than in the single-observation case.  This
demonstrates that performing the measurement in two independent load
cases effectively over-determines the inverse problem, in turn
yielding an estimation with narrower parameter uncertainty ranges.

As a complement to the estimated model parameters, the ratio of
characteristic lengths can be obtained from their sampled values.  For
the single observation, the MAP estimate is $\Lambda'/\Lambda=2.241$
and the 95\% credible interval is $\Lambda'/\Lambda\in[2.122,2.326]$.
The dual observation yields a MAP estimate of
$\Lambda'/\Lambda=2.246$, and a narrower 95\% credible interval,
$\Lambda'/\Lambda\in[2.162,2.316]$.  It is worth noting that the
results with both observation approaches respect the well-established
condition $\Lambda'/\Lambda\geq2$~\cite{AllardAtalla2009}.

In addition to the poroelastic material properties, the dimensions of
the expansion chamber are obtained as a by-product of the method and
can serve as part of the validation.  In the present work we have
chosen to introduce these geometrical properties as additional
unknowns, but alternatively they could be assumed at their nominal
values, albeit at the risk of introducing a bias in the values of the
properties of the sample.  Indeed, the present analysis shows that the
geometrical parameters of the expansion chamber and coupling tube
differ from their nominal values.  For instance, the radius $\rexp$ of
the expansion chamber presents an apparent overestimation.  In the
opinion of the authors this arises as a compensation of one or both of
the limitations of the model in this respect.  The first possible
cause is the well-known intrinsic inability of the classical transfer
matrix method to correctly represent three-dimensional effects at
sudden
expansions~\cite{Munjal1987,AlkmimCuencaDeryckGoransson2018ISMA}, and
the second is the unaccounted-for finite stiffness of the expansion
chamber.  It is worth noting that the uncertainty ranges in the manual
measurements are directly derived using the precision of the length
readings.  This is naturally different from the statistical definition
of 95\% credible intervals, which prevents a rigorous comparison.
Nevertheless, the measurements of lengths $h$, $\lb$, $\lexp$ and
$\lc$ overlap with the model inversion results.  In particular, the
length correction procedure correctly predicts the length of the
coupling tube $\lc$ and that of the primary tube section $\lb$.

\section{Summary of the findings}\label{sec:discussion}
Based on the results above, the primary findings of this work can be
summarised as follows.  First, the proposed transfer matrix model of
the system provides an accurate representation of the behaviour of the
material sample under both loading conditions \loadA\ and \loadB, thus
enabling a good fit onto the experimental acquisitions while
exhibiting sufficient sensitivity to all model parameters.

The proposed incremental inversion method has been shown to guarantee
that the global solution of the problem is obtained independently from
the initial guess, albeit at a significant computational cost as
compared to a full-range search.  This has been illustrated for both a
single observation in load case \loadA\ and a dual observation
combining load cases \loadA\ and \loadB.

Furthermore, the proposed framework for statistical model inversion
provides marginal posterior densities for all unknowns of the problem.
Accordingly, this enables a reliable extraction of point estimates and
credible intervals.  A notable demonstration of the robustness of the
method is that the credible intervals for the single- and
dual-observation cases overlap.  More specifically, the point
estimates for the single-observation case are systematically contained
in the credible intervals of the dual-observation case, and vice
versa.

The central result of this manuscript is the observation that the
credible intervals are narrower in the case of the dual observation.
This shows that the use of a multi-observation dataset provides an
over-determination of the inverse problem, thereby yielding refined
uncertainty ranges on the unknowns.

A triple-observation model inversion was also performed, using an
additional load case with a coupling tube of nominal length
$\lcO=30\,$mm.  However, the results were comparable to the
dual-observation case and no improvement was observed, presumably due
to the fact that the additional load case did not significantly
over-determine the inverse problem.  These results are not reported
here.

The application example shows that the method is capable of providing
the full set of properties required to model
macroscopically-homogeneous poroelastic media under the
Biot-Johnson-Champoux-Allard model, with the exception of the Poisson
ratio, from two-microphone impedance tube measurements.  In the
opinion of the authors, this achievement is of substantial value as
previous methods rely on the assumption that one or more of these
parameters are initially known, therefore removing the need for
multiple test rigs.

Finally, in addition to point estimates and credible intervals, the
statistical inversion framework allows for correlations among the
parameters to be observed.  Using the present results, these can be
tied to a certain extent to the one-degree-of-freedom scaling laws for
poroelastic
media~\cite{GibsonAshby1999,Goransson2006,LindnordgrenGoransson2010,CameronLindnordgrenWennhageGoransson2014}.

\section{Conclusion}\label{sec:conclusion}
A multi-observation approach is here proposed, based on two
consecutive measurements of the sound absorption coefficient of a
rigid-backed sample of a poroelastic material, where in the second the
sample is coupled to an expansion chamber.  The methodology is shown
to provide all 9 parameters of a comprehensive poroelastic material
model, including the sample thickness and density, 5 transport
parameters and 2 elastic properties.  This is achieved through a
combination of an incremental method designed to increase the
complexity of a deterministic inverse problem gradually, and a
statistical inversion framework providing the posterior probability
density of the model parameters, from which point estimates, credible
intervals and correlations are inferred.  The results show reasonable
credible intervals for all material parameters in accordance with
values reported for melamine in the literature.  Furthermore, the
over-determination induced by the use of a multi-observation approach
is shown to reduce the uncertainty in the parameter estimation.
Additionally, the correlations observed among the porous material
properties support the claim that the classical parametrisation of the
acoustical behaviour of open-cell porous media is somewhat redundant
and can be replaced with a model with fewer
parameters~\cite{HoroshenkovHurrellGroby2019jasa}.

The method has been here illustrated for a melamine foam sample, for
which the dimensions of the expansion chamber and coupling tube were
suitable for enhancing fluid-structure interaction with the frame's
elasticity.  Applying the method to other materials may require
different dimensions of the expansion chamber loading in order to
promote a suitable degree of coupling.  In future work, further
over-determining the inverse problem could be achieved for instance by
varying the expansion chamber length, which controls the major
resonances of the coupled system.

In light of the results obtained by Niskanen et
al.~\cite{NiskanenGrobyDuclosDazelLerouxPoulainHuttunenLahivaara2017},
the present method could be further applied to measurements in a
three-microphone impedance tube with a rigid-backed sample, for
instance using the processing proposed in
Refs.~\cite{IwaseIzumiKawabata1998,SalissouPannetonDoutres2012JASAEL}.
This would enable the additional measurement of the equivalent density
and bulk modulus, while ensuring a proper control of the boundary
conditions.

It is worth noting that the proposed incremental inversion procedure
is applicable to a wider variety of problems so long as modelling
complexity can be treated gradually, for instance exploiting an
asymptotic behaviour in high frequencies, or along other conditional
variables of interest such as temperature or mechanical stress, for
example.

\section*{Acknowledgements}
This article is based upon work initiated under the support from COST
Action DENORMS CA-15125, funded by COST (European Cooperation in
Science and Technology).  The work has been supported by the Academy
of Finland (Finnish Centre of Excellence of Inverse Modelling and
Imaging and project number 321761) and by the Centre for ECO2 Vehicle
Design at KTH, Vinnova Grant No. 2016-05195.

\end{document}